\documentclass[aps,pra,twocolumn,superscriptaddress,showpacs]{revtex4}

\usepackage{graphicx}
\usepackage{times}
\usepackage{color}
\usepackage{amsmath,amssymb}

\makeatletter
\renewcommand{\Dated@name}{Date: }
\makeatother

\newcommand{\ket}[1]{\mbox{$ | #1 \rangle $}}
\newcommand{\bra}[1]{\mbox{$ \langle #1 | $}}
\newcommand{\tpaper}{arXiv:1310.8465}
\newcommand{\rhoLIN}{{\hat \rho_{\mathrm{LIN}}^{\ }}}
\newcommand{\rhoMLE}{\hat \rho_{\mathrm{MLE}}^{\ }}
\newcommand{\rhoFLS}{\hat \rho_{\mathrm{FLS}}^{\ }}
\newcommand{\rhoT}{{\rho_{\textsc{t}}^{\ }}}
\newcommand{\tr}[1]{\textrm{Tr}{\left\{#1\right\}}}
\newcommand{\barr}[1]{\overline{#1}}
\newcommand{\mathvisiblespace}{\mbox{\textvisiblespace}}

\begin{document}

\title{Quantum state tomography: Mean squared error matters, bias does not}
\date[]{Dated: \today}

\author{Jiangwei Shang}
\affiliation{Centre for Quantum Technologies, National University of Singapore, %
3 Science Drive 2, Singapore 117543, Singapore}

\author{Hui Khoon Ng}
\affiliation{Centre for Quantum Technologies, National University of Singapore, %
3 Science Drive 2, Singapore 117543, Singapore}
\affiliation{Yale-NUS College, 6 College Avenue East, Singapore 138614, Singapore}

\author{Berthold-Georg Englert}
\affiliation{Centre for Quantum Technologies, National University of Singapore, %
3 Science Drive 2, Singapore 117543, Singapore}
\affiliation{Department of Physics, National University of Singapore, %
2 Science Drive 3, Singapore 117542, Singapore}

\begin{abstract}
Because of the constraint that the estimators be \textsl{bona fide}
physical states, any quantum state tomography scheme---including the widely
used maximum likelihood estimation---%
yields estimators that may have a bias, although they are consistent estimators.
Schwemmer \textsl{et al.} (arXiv:1310.8465 [quant-ph]) illustrate this by
observing a systematic underestimation of the fidelity and an overestimation
of entanglement in estimators obtained from simulated data.
Further, these authors argue that the simple method of linear inversion
overcomes this (perceived) problem of bias, and there is the suggestion to abandon
time-tested estimation procedures in favor of linear inversion.
Here, we discuss the pros and cons of using biased and unbiased estimators
for quantum state tomography.
We conclude that the little occasional benefit from the unbiased
linear-inversion estimation does not justify the high price of using
unphysical estimators, which are typically the case in that scheme.
\end{abstract}

\begin{widetext}
\maketitle
\end{widetext}

\section{Introduction}
The goal of quantum state tomography, or quantum state estimation, is to
arrive at a best guess of the unknown state $\rho$ of a quantum system, based
on data collected from measuring a number of identical copies of the state.
An accurate guess is needed in all aspects of quantum information or quantum
computation, ranging from the characterization of an unknown quantum
communication channel, to a check of a quantum gate implementation, or to the
verification of a state preparation procedure in the lab.

Whether the guess from a particular tomography recipe can be considered the
best, or most accurate, depends on one's figure-of-merit, which should be
chosen according to the quantum information processing task at hand.
In many situations, one is interested not in the state of the system itself,
but in a quantity computed from it, \textsl{e.g.}, the amount of entanglement
in the state.
In such cases, rather than reporting a best guess for $\rho$, one expects to
get a more accurate answer by directly estimating the quantity of interest
from the data, as done in a related procedure carrying the name of ``parameter
estimation.''
In other situations, one is interested in a range of quantities related to
$\rho$, and reporting a best guess for the state itself [\textsl{e.g.}, by
maximizing the likelihood for the data over all physical states, as is done
for the maximum-likelihood estimator (MLE)] can be a convenient and
self-consistent way of summarizing and interpreting the data.

In the latter case, one might assess the accuracy of the estimate obtained
from a particular tomography scheme by examining some measure of closeness
between the estimate $\hat\rho$, and the true state $\rho$, for a variety of
known true states (\textsl{e.g.}, from sources that have previously been fully
characterized).
A poorer, but possibly useful, gauge is to look at the accuracy of the prediction
of one of the quantities of interest computed from $\hat\rho$.

Reference~\cite{\tpaper} compares the performance between an estimator
$\rhoLIN$ from a procedure the authors refer to as linear inversion (LIN),
and two other estimators, $\rhoMLE$ (the standard MLE \cite{ML1997,LNP649}),
and $\rhoFLS$ (from a procedure known as ``free least squares'' (FLS) \cite{FLS}).
The article assesses the three estimators by looking at the accuracy in the
prediction for target fidelity, a relevant quantity when the source is
supposed to produce a certain target state.

The fidelity $F(\rho_1^{\ },\rho_2^{\ })\equiv%
\tr{\biglb|\sqrt{\rho_1^{\ }}\sqrt{\rho_2^{\ }}\,\bigrb|}^2$ takes value
between $0$ and $1$ for physical $\rho_1^{\ }$ and $\rho_2^{\ }$
\cite{note:FvsFsquare}.
In Ref.~\cite{\tpaper}, the quality of an estimator is measured by the target
fidelity $F(\rhoT,\hat\rho)$, the fidelity between the target state $\rhoT$ and
the estimator $\hat\rho$, which is compared with the actual true value
$F_0\equiv F(\rhoT,\rho)$ computed for the true state $\rho$.
Here, the ``true'' state yields the probabilities that are used for the
generation of the simulated data from which the various estimators are derived.

The authors of Ref.~\cite{\tpaper} perform this comparison for different
$\rhoT$s and $\rho$s, for many repeated simulations of the measurement data
(and hence, many $\hat\rho$s, one for each data).
They draw the conclusion that $\rhoLIN$ is always the best, because it is
an unbiased estimator, \textsl{i.e.}, fluctuations from different runs of the
same experiment lead to fluctuations in the predicted $F$ value, but all
centered about the true value $F_0$; $\rhoMLE$ and $\rhoFLS$, on the other
hand, give predictions that are biased, \textsl{i.e.}, have a systematic
shift away from $F_0$ (see Fig.~1 of Ref.~\cite{\tpaper} and
Fig.~\ref{fig:NoisyGHZ} below).
The authors go on to point out that any estimation procedure that always
produces a physical (\textsl{i.e.}, nonnegative) state will unavoidably
be biased; their $\rhoLIN$, coming from an unbiased estimation procedure, is
not guaranteed to be a physical density operator, and in fact, generically has
negative eigenvalues \cite{\tpaper}.

Here, the qualifiers \emph{biased} and \emph{unbiased} have the technical meaning
that is discussed below in the context of Eq.~(\ref{twopieces}).
Contrary to their connotation in common parlance, they are not synonyms of
``bad'' and ``good.''
One must not fall into the trap of regarding a biased estimator as
automatically inferior to an unbiased one.

Indeed, it is well known in classical statistics that
unbiased estimators are not always the best choice.
Instead, minimizing the mean squared error (MSE, a popular measure of
estimation accuracy) is key,
and this is often not accomplished by minimizing the bias.
In fact, we will show that the LIN approach yields MSEs that
are comparable to (and sometimes worse than) what one obtains from the MLE
procedure; yet the LIN technique forces us to give up physicality, which leads
to many severe problems and highly restricts the usefulness of the estimator
$\rhoLIN$.
The MLE itself also does not---and was never designed to---minimize the MSE,
but it does a comparably good job as $\rhoLIN$ while enforcing
physicality.
We hence see little utility at all in employing the LIN strategy.

Below, we remind the reader why, in the quantum context, it is usually not a
good idea to treat relative frequencies obtained from the data as
probabilities, as is prescribed by the LIN procedure of Ref.~\cite{\tpaper}.
Then, we explain why focusing on reducing bias only, and not the overall MSE,
constitutes a conceptual misunderstanding.
Lastly, we compare the MSEs obtained from the MLE and the LIN approaches
and observe that one can easily find examples in which the biased MLEs
have smaller MSEs than the unbiased LIN estimators.

\section{Frequencies are not probabilities}

Before we begin, a brief note on notation is in order.
The tomography measurement is described by a positive-operator-valued
measure (POVM), or, if we use a more descriptive name, a probability-operator
measurement (POM):
It comprises a set of outcomes $M_k$, one for each detector, with $M_k\geq 0$
for all $k$ and $\sum_k M_k=1$.
The probability of getting a click in detector $k$, corresponding to outcome
$M_k$, is given by the Born rule, $p_k=\tr{\rho M_k}$.
In the tomography experiment, $N$ identically prepared copies of the (unknown)
state $\rho$ are measured using the POM.
The data $D$ consist of a particular sequence of detector clicks, summarized
by the set of relative frequencies $f_k=n_k/N$, where $n_k$ is the number of
clicks in detector $k$.
From $D$, one estimates the probabilities $\hat p_k$ using a chosen procedure
like MLE or LIN, and from these (if one can, \textsl{e.g.}, in the case of
tomographically complete POMs), one constructs the estimator $\hat\rho$.

LIN, as proposed in Ref.~\cite{\tpaper}, sets the estimated probabilities
equal to the relative frequencies of the observed data,
\begin{equation}
\hat p_k=f_k\,,
\end{equation}
and then obtains $\rhoLIN$ by ``linear inversion'' of the Born rule
${\hat{p}_k=\tr{\rhoLIN M_k}}$.
While relative frequencies will be close to probabilities when there is a lot
of data, they are most certainly not the same thing:
Relative frequencies satisfy only one constraint, that of unit sum: $\sum_kf_k=1$;
probabilities $p_k$ (for POM $\{M_k\}$) that arise from a physical state
$\rho$ through the Born rule satisfy further constraints imposed by the
positivity of $\rho$.
The latter constraints can be easily stated for the case of measuring a qubit
state with the symmetric informationally complete POM (SIC POM), the
tetrahedron measurement \cite{PRA70.052321,JMP45.2171}, where $\rho\geq 0$
requires the four tetrahedron probabilities to satisfy
$\sum_kp_k^2\leq \frac{1}{3}$, in addition to the unit-sum constraint.
Since the relative frequencies do not themselves satisfy these physicality
constraints, $\rhoLIN$ is hence not necessarily a physical state, as is also
emphasized in Ref.~\cite{\tpaper} (and many other existing references in the
literature).

That $\rhoLIN$ is not necessarily nonnegative, is not a minor nuisance:
Many quantities associated with a physical state $\rho$ are ill-defined for
$\rhoLIN\not\geq0$ and can no longer be computed, \textsl{e.g.},
entropy, negativity, and the fidelity with another state.
Other quantities, such as the purity $\tr{{\rhoLIN\,\!}^2}$ or the expectation value
$\tr{A\rhoLIN}$ of an observable $A$, are computable for $\rhoLIN\not\geq0$,
but the numbers so obtained do not mean purity, expectation value, etc.
Hence, $\rhoLIN$ may not just lack a reasonable physical interpretation, but
may also not be useful at all.
In the case of uncomputable quantities, the proposal of Ref.~\cite{\tpaper}
is to be content with the bounds that can be computed from linear approximations.
These bounds, however, also lack a physical meaning if they are evaluated
for $\rhoLIN\not\geq0$.

While one might choose not to be too concerned if $\rhoLIN$ is only slightly
unphysical (however one may want to quantify that statement), or if an
unphysical $\rhoLIN$ occurs only rarely, getting an unphysical $\rhoLIN$ can
be generic in certain situations.
For example, imagine a qubit state measured with the tetrahedron measurement,
and suppose that the true state is orthogonal to one of the tetrahedron
outcomes (say the one labeled by $k=1$).
Then, the only relative frequencies that can give a physical $\rhoLIN$
are $f_1=0$ and ${f_2=f_3=f_4=\tfrac{1}{3}}$, \textsl{i.e.}, the
detector counts for all outcomes, other than the tetrahedron leg orthogonal to
the true state, must be exactly equal.
This is not even possible if the total number of counts is different from a
multiple of 3.

\begin{figure}
\includegraphics[width=0.95\columnwidth,trim = 50mm 210mm 80mm 40mm,clip]%
{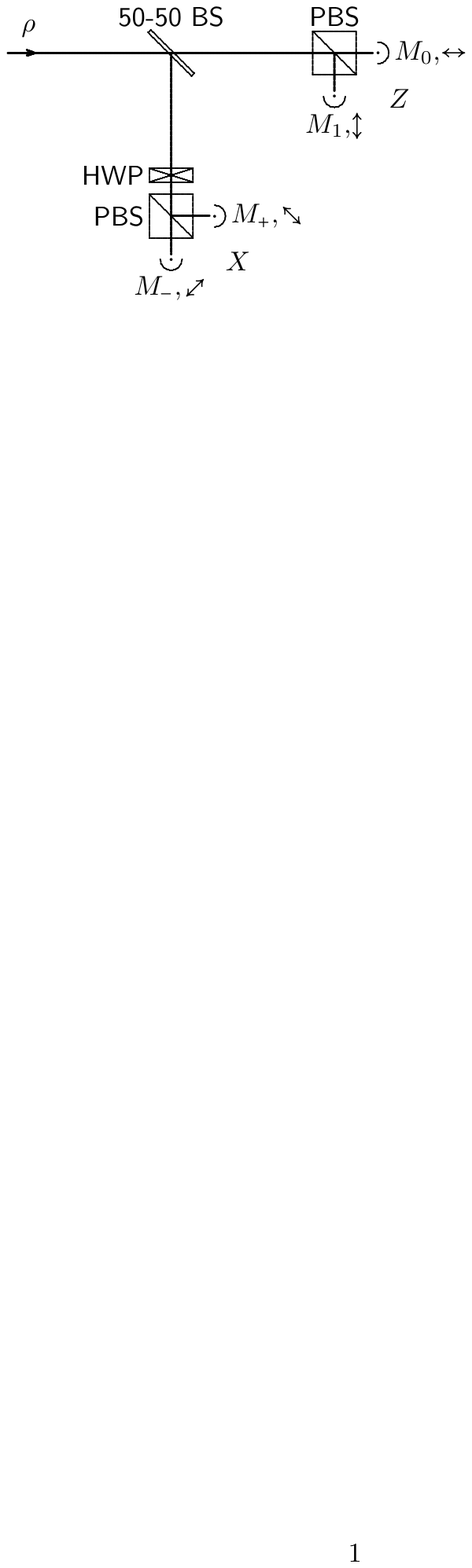}
\caption{\label{fig:BB84} $X$ and $Z$ measurements of a single qubit.
The set-up uses a beam splitter (BS), a half-wave plate (HWP), two polarizing
beam splitters (PBSs), and four detectors.}
\end{figure}

Lest the reader complains that the above is a pathological case, another
situation where one sees a stark contrast between frequencies and
probabilities can be found in the commonly used ``BB84-like'' measurements,
\textsl{i.e.}, measure the Pauli $X$ and $Z$ on a single qubit with equal
probability.
In an optical implementation, where the qubit is the photon polarization, the
usual way this measurement is implemented is by having a 50-50 beam-splitter
direct the incoming photons into two possible paths, one carrying out
the $X$ measurement, the other the $Z$ measurement; see Fig.~\ref{fig:BB84}.
Now, the probabilities for such a measurement, by the very nature of the
measurement structure, satisfy ${p_0+p_1=p_++p_-=\frac{1}{2}}$, alongside
the positivity constraint ${(p_0-p_1)^2+(p_+-p_-)^2\leq\frac{1}{4}}$.
The relative frequencies, however, obey no such constraints:
Despite the 50-50 nature of the beam-splitter, one hardly ever encounters the
situation where \emph{exactly} half the photons travel down the $X$ path, and
half down the other.
The procedure of finding $\rhoLIN$ from such relative frequencies will then
typically be internally inconsistent, and yields no solution.

One common fix used to circumvent the above problem
requires one to ignore the counts in one of the detectors, \textsl{e.g.}, the
one measuring the $+1$ eigenstate for $X$
(outcome $M_+$ in Fig.~\ref{fig:BB84}) \cite{FLS}.
To ensure that the relative frequencies comply with the constraint of
${f_0+f_1}={f_++f_-}=\frac{1}{2}$, in imitation of the probabilities, one
replaces the actual count $n_+$ obtained by ${n_+=n_0+n_1-n_-}$, and modifies
the total number $N$ to be $2(n_0+n_1)$.
However, this ad-hockery, which involves discarding data, does not
guarantee that the relative frequencies satisfy the remaining positivity
constraint on the probabilities.

The simulations in Ref.~\cite{\tpaper} mimic the tomography of four-qubit
states using product Pauli POMs, \textsl{i.e.}, measure
${[O_{1}O_{2}O_{3}O_{4}]\equiv O_{1}\otimes O_{2}\otimes O_{3}\otimes O_{4}}$, with
each of the four $O_{\alpha}$s equal to one of the three Pauli operators $X$,
$Y$, and $Z$.
Rather than just having two settings of $X$ and $Z$ as in the single-qubit case
of Fig.~\ref{fig:BB84}, there are now ${3^4=81}$ settings, all to be measured
with equal probability.
A practical way of carrying out this tomography measurement, having (usually)
no access to a 1-to-81 beam-splitter, would be to divide the total number of
counts by $81$, and measure $N/81$ copies of $\rho$ with each setting (this is
another way of overcoming, by hand, the difficulty discussed in the previous
two paragraphs).
This automatically ensures that the relative frequencies for each setting
$[O_{1}O_{2}O_{3}O_{4}]$ sum to $1/81$, as do the corresponding probabilities.

However, these constraints are not the only ones needed to ensure internal
consistency.
There is, for example, the problem that frequencies obtained when
measuring, say, $[XZZY]$ and $[XYXY]$ give two values for the expectation values
of the two-qubit observable
$[X\mathvisiblespace\mathvisiblespace Y]=X\otimes1\otimes1\otimes Y$ and
also for the single-qubit observables
$[X\mathvisiblespace\mathvisiblespace\mathvisiblespace]$
and $[\mathvisiblespace\mathvisiblespace\mathvisiblespace Y]$.
In Ref.~\cite{\tpaper}, these conflicts are resolved by a fix that may
appear plausible, but is ad-hockery nevertheless and ultimately difficult to
justify.
Furthermore, there is still the issue of positivity constraints, and our own
simulations show that almost all the estimators constructed by LIN violate
positivity and, therefore, are unphysical.

None of these ad-hoc fixes or discard of data are necessary if we do
not insist on setting frequencies equal to probabilities.
The usual approach of finding a \emph{physical} density operator that best
fits the data, \textsl{e.g.}, by maximizing the likelihood for the data over
all physical states in the case of the MLE, automatically handles all these
constraints.
There is no need to include constraints explicitly by hand, such a need
becoming more severe and difficult to carry out as the dimensionality of the
system and the complexity of the POM increase.
Of course, finding the best-fit physical estimator for the data is also not
easy, but the estimator is assuredly physical, as is the true state, and one
interprets the data honestly without additional tweaks.

\vfill

\section{To bias or not to bias?}
The accuracy of an estimator is often quantified by the mean-squared error
(MSE), \textsl{i.e.}, the mean, over all possible data, of the squared
deviation of the estimated parameter from the true value, but this is
only one of many measures of inaccuracy.
To be concrete, we take as example the estimate $\hat F$ for the target
fidelity (with respect to some target state $\rhoT$) as discussed in
Ref.~\cite{\tpaper}.
The MSE in this case is
\begin{equation}\label{twopieces}
  \mathbb{E}_{\rhoT}[(\hat{F}-F_0)^2]=[\mathbb{E}_\rhoT(\hat{F})-F_0]^2
 +\mathbb{V}_\rhoT(\hat{F})\,,
\end{equation}
where we adopt the notation of Ref.~\cite{\tpaper}:
$\mathbb{E}_\rhoT$ denotes the expected value and $\mathbb{V}_\rhoT$ the
variance, for the true state $\rhoT$.
The MSE in Eq.~(\ref{twopieces}) is the sum of two contributions:
The first is the so-called \emph{bias} of the estimator
(as compared with the true value $F_0$);
the second is the variance of $\hat F$.
As it is written, the formula for the MSE makes no statement about the
relative importance of the two pieces, and a general estimation strategy can
have different relative sizes for, and compromises between, the two
terms.
Other measures of inaccuracy may not have a break-up analogous to
Eq.~(\ref{twopieces}), and the bias may not be a relevant notion.

It is easy to understand intuitively why both pieces, not just the bias, must
be small in order for us to have reasonable confidence of obtaining a good
estimate from a single set of data. If the bias is small, and one has many
runs of the same experiment, yielding very many estimates, one might imagine
getting a good answer by looking for the centre of all these
estimates. However, one usually has access to only one set of data, and hence
a single estimate \cite{note:data-split}; other estimates may then be
generated by bootstrapping the data, but this just gives a distribution of values
centered at the data-based estimate.
If the estimation strategy yields no bias, but a large variance, one will
often end up with an estimate that is quite far away from the true
value. Hence, one needs both the bias, as well as the variance, summarized as
the MSE, to be small for an accurate guess.

Most statistics textbooks (see, for example, \cite{Lehmann}) do focus on the
class of unbiased estimators for a given problem.
However, this is always discussed together with the search for one with the
minimum variance, \textsl{i.e.}, to minimize the MSE \emph{within the class of
  unbiased estimators}.
The resulting theory of point estimation for unbiased estimators has rather
nice and concise results, with wide applicability in many areas of
statistics.
Yet, at the same point where unbiased estimators are discussed, textbooks
usually remind readers that unbiased estimators are not the whole story.
The constraint of unbiasedness is often too strong: An estimator with a slight
bias but a significantly smaller variance can very well be the estimator that
minimizes the MSE.
Certainly, considering only unbiasedness without also trying to minimize the
variance, as suggested in Ref.~\cite{\tpaper}, is altogether insufficient to
give a good estimator.

Furthermore, as most textbooks will also explain, reasonable unbiased
estimators do not exist in many problems.
The nonexistence of an unbiased estimator that satisfies the positivity
constraint for quantum states, as is emphasized in Ref.~\cite{\tpaper}, is but
another example of the difficulty in requiring unbiasedness in constrained
estimation problems.
As a result of constraints, pure states and other rank-deficient states are extremal and the
valid (\textsl{i.e.}, satisfy the constraints) estimators cannot approach them
from all sides: The estimators are unavoidably biased.
This is clearly visible in the examples discussed in the next section.

The trustworthy estimators, such as the MLEs, are \emph{consistent},
\textsl{i.e.}, they converge to the true state when more and more data are
taken into account.
This consistency is an important properties that is worth insisting upon,
while unbiasedness is not.

The various virtues of, and problems with, unbiased estimators
have been repeatedly expounded upon in the statistics literature.
As an example, we point the reader to the excellent discussion found in
Jaynes's book on probability theory, Ref.~\cite{JaynesBook} (see Chapter 17).
Within Jaynes's book is an example where the requirement of unbiased
estimators leads one to waste half the data as compared with
a biased estimator that attains the same MSE \cite{note:irony}.

\section{MSE for LIN and MLE approaches}

To illustrate the issues mentioned in the previous sections, we re-examine the
example of target fidelity discussed in Ref.~\cite{\tpaper}. The unknown true
state is a four-qubit state, and the parameter of interest is its fidelity
with some target state $\rhoT$.
The measurement comprises the $81$ product Pauli operators described above,
with $100$ copies measured per setting, and $16$ possible outcomes per
setting.
To explore the statistics, for each set of target and true states, we simulate
$500$ runs of the experiment, obtain the estimators $\rhoMLE$ and $\rhoLIN$
for each set of data, and from these, estimate the target fidelity.

As observed above, there is a problem with evaluating (not to mention,
interpreting the meaning of) the target fidelity
$F(\rhoT,\rhoLIN)=\tr{\biglb|\sqrt{\rhoT}\sqrt{\rhoLIN}\,\bigrb|}^2$
when $\rhoLIN$ is not nonnegative.
Reference~\cite{\tpaper} also admits this problem, and suggests to fix it by
using the
formula ${F(\rhoT,\rhoLIN)=\tr{\rhoT\rhoLIN}}$ instead, which is correct \emph{if}
$\rhoLIN$ is nonnegative and $\rhoT$ is a pure state.
Although $\tr{\rhoT\rhoLIN}$ can be computed also for an unphysical $\rhoLIN$,
it does not have the significance of the target fidelity then.
The other issue, \textsl{viz.} that $\rhoT$ must be pure, is of lesser concern
if all target states of interest are indeed pure; the Smolin state considered
in Ref.~\cite{\tpaper} is not pure, and this is why the respective line
for $\rhoLIN$ is missing in their Fig.~2.
Clearly, one already sees the problems with using a nonphysical estimator.

\begin{figure}
\includegraphics[width=0.95\columnwidth]{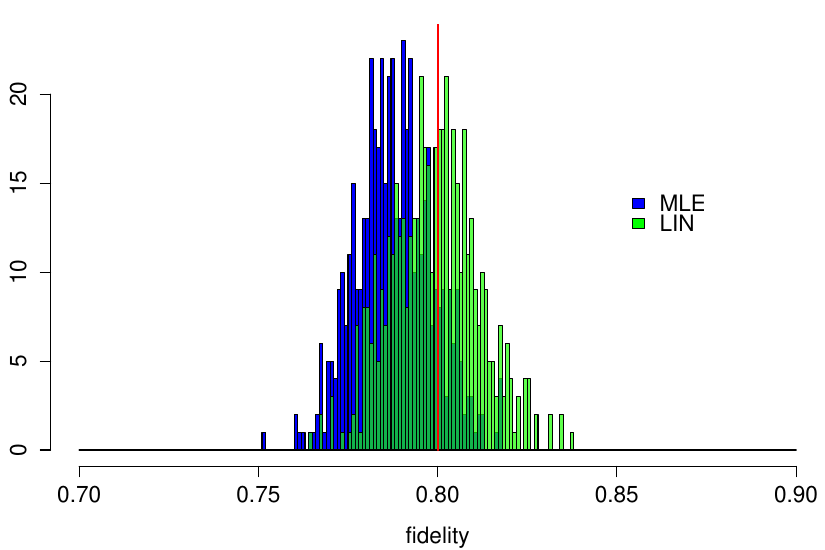}
\caption{\label{fig:NoisyGHZ}%
Fidelity histograms of simulated quantum-state tomography experiments
for the noisy GHZ state from Ref.~\cite{\tpaper}.
The true state is a GHZ state with added white noise so that its fidelity is
$0.8$ with respect to the original target GHZ state.}
\end{figure}

Nevertheless, we can still compare the estimated target fidelities obtained
from the MLE and LIN procedures for pure target states only, using their
suggested formula of
$F(\rhoT,\hat\rho)=\langle\psi|\hat\rho|\psi\rangle$ for
$\rhoT=\ket{\psi}\bra{\psi}$, whether $\hat\rho$ is physical or not.
As the first target state, we consider the four-qubit
Greenberger-Horne-Zeilinger (GHZ) state with the ket
$\ket{\mbox{GHZ}_4}=\tfrac{1}{\sqrt 2}(\ket{0000}+\ket{1111})$.
This is the main example discussed in Ref.~\cite{\tpaper}; we simulate
the identical situation and the results are shown in Fig.~\ref{fig:NoisyGHZ},
which are very similar to those in Fig.~1 of Ref.~\cite{\tpaper}.

\begin{figure}
\includegraphics[width=0.95\columnwidth]{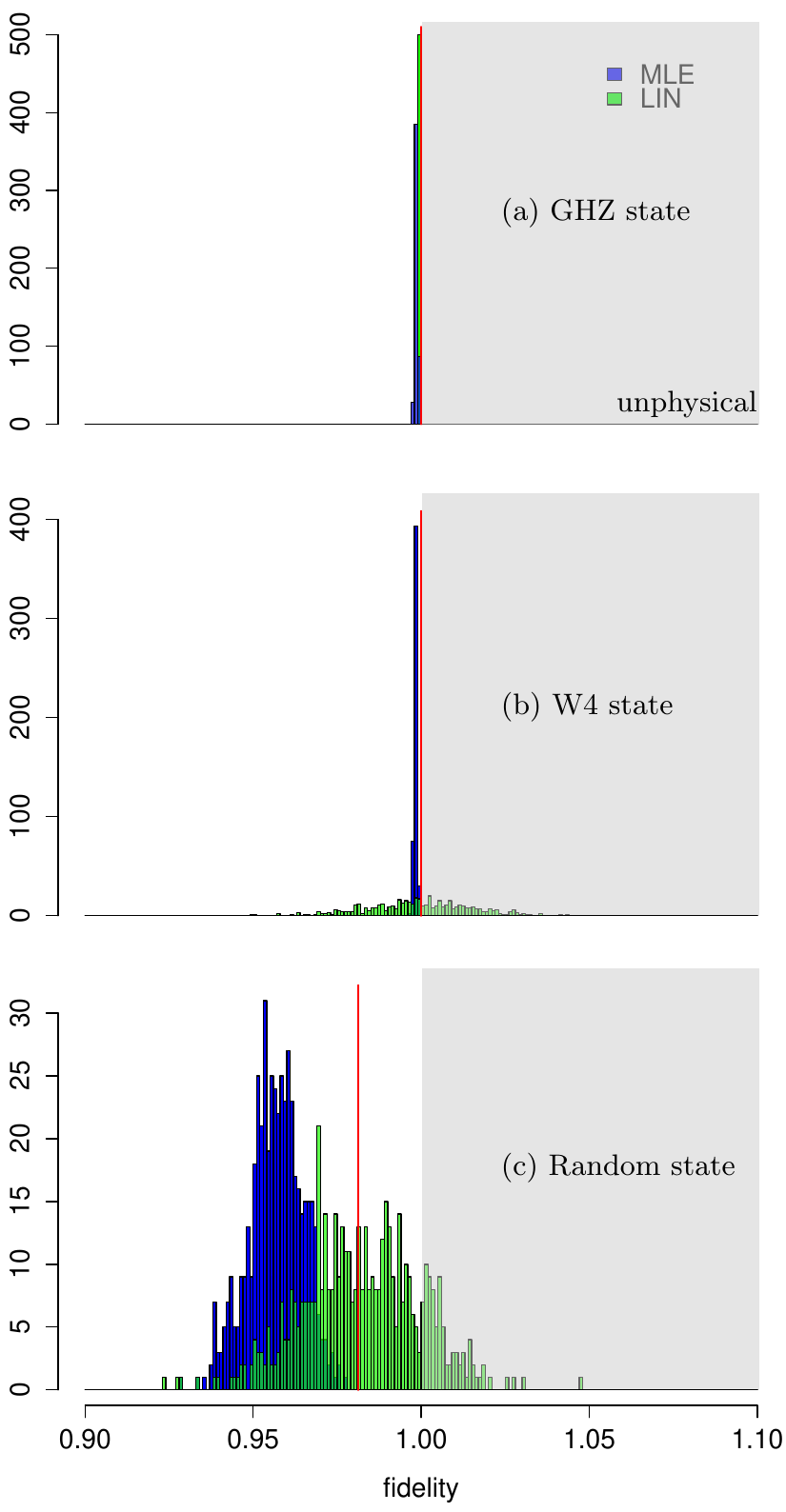}
\caption{\label{fig:Plots}%
Fidelity histograms of simulated quantum-state tomography experiments
for various states (indicated in plots).
The GHZ and W4 states of plots (a) and (b) have true fidelity of $1$;
plot (c) is for a true state that has fidelity $0.981$ with a
randomly chosen pure target state.
In plots (b) and (c), the histograms for the LIN values extend into the
unphysical range of fidelity values exceeding unity.}
\end{figure}

Instead of mixing the GHZ state with white noise, we can also take the GHZ
state itself as the true state. Then the true fidelity with respect to the
target state takes the maximal value of $1$.
In this situation, we find that all the fidelity values estimated using LIN
give the true value of $1$, while the MLE values slightly underestimate the fidelity
(see Fig.~\ref{fig:Plots}(a); $\barr{F}_{\textsc{MLE}}=0.999\pm0.00036$, the
mean and the standard deviation obtained from 500 data sets).
That LIN always gives the correct fidelity value, regardless of the
statistical fluctuations in the data, is a special feature of the GHZ target
state in conjunction with the product Pauli POM \cite{note:GHZsingular}.
Such a singular situation (as can be seen from more generic
examples below) does not provide a fair benchmark from which to draw
reliable conclusions about the efficacy of the estimation procedures.

As the second example, we use a different target state, also
taken from Ref.~\cite{\tpaper}, namely the four-qubit W-state,
$\ket{\mbox{W}_4}=\tfrac{1}{2}(\ket{0001}+\ket{0010}+\ket{0100}+\ket{1000})$.
To illustrate the problems of a nonphysical $\rhoLIN$, we again consider the
extreme scenario where the true state is exactly the target state so that the
true fidelity is $1$.
LIN gives fidelity values ($\barr{F}_{\textsc{LIN}}=0.999\pm0.015$) that are
distributed
around the true value but have a wide spread, with a large fraction of the
values exceeding $1$, a situation without physical meaning.
The results obtained via MLE ($\barr{F}_{\textsc{MLE}}=0.998\pm0.00042$) again slightly
underestimate the fidelity, but not by much, and the spread of values is
much narrower than for LIN; see Fig.~\ref{fig:Plots}(b).
That the MLE underestimates the true fidelity value is unavoidable because
$F\leq 1$ for physical states and this demonstrates our earlier point that one
can only approach an extremal value from one side.

The GHZ and W states considered thus far are quite special states, and
particularly so with respect to the product Pauli POM.
One wonders about the behaviour for a generic target state.
To this end, we randomly pick a pure target state $\ket{\phi}$ and a random
true state $\rho$ with a target fidelity close to $1$.
The selected true state has target fidelity of $0.981$; the fidelities
calculated via LIN give $\barr{F}_{\textsc{LIN}}=0.982\pm0.018$ and those
gotten via MLE give $\barr{F}_{\textsc{MLE}}=0.957\pm0.0080$; see
Fig.~\ref{fig:Plots}(c).
Again, a significant fraction of fidelity values obtained from LIN are greater
than $1$; no such problems appear in the MLE case, by construction, and the
MLE range of values has a smaller spread.

\begin{figure}
\includegraphics[width=0.95\columnwidth]{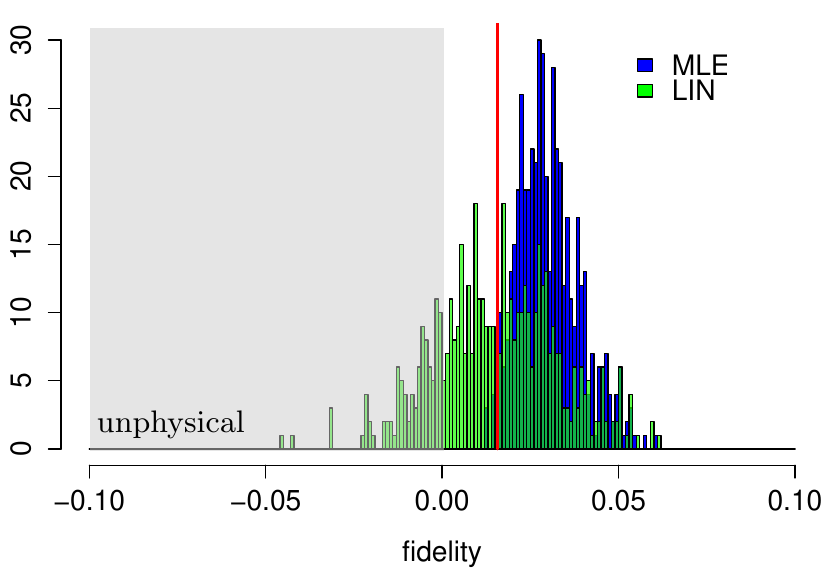}
\caption{\label{fig:rand_ort} Fidelity histograms of simulated quantum-state
  tomography experiments
  for the same noisy random pure state as in Fig.~\ref{fig:Plots}(c), but with
  the true fidelity being $0.016$.
  The histograms for the LIN values extend into the unphysical range of
  negative fidelity values.}
\end{figure}

The same problems of unphysical fidelity values appear at the other end
of the allowed fidelity range, \textsl{i.e.}, for $F$ close to $0$.
We again use the same randomly generated pure state $\ket{\phi}$ as the
target state, but now choose a true state with a target fidelity of $0.016$.
This is the situation in which one's intent is to prepare a state
orthogonal to the target state.
In this setting, we find that the fidelities calculated via LIN
($\barr{F}_{\textsc{LIN}}=0.015\pm0.017$) still center around the true value,
but with a large fraction being less than $0$, another situation that is
physically meaningless.
On the other hand, the values obtained via MLE
($\barr{F}_{\textsc{MLE}}=0.030\pm0.0088$) in this case slightly overestimate
the fidelity, but again the spread of MLE values is smaller than the
spread of LIN values; see Fig.~\ref{fig:rand_ort}.

\begin{table*}
\caption{\label{tbl:MSE}
The MSE as well as the respective variance and bias in different examples with
pure target states.
The first column states the ket of the four-qubit target state used
in the simulation,
with respect to which the true state has the fidelity $F_0$ of the second column.
The kets $\ket{\phi}$, $\ket{\gamma}$, $\ket{\tau}$, and $\ket{\theta}$ are
randomly chosen examples.
In the third, fourth, and fifth columns, the double entries report the LIN
value on the left of the $|$ symbol and the MLE value on the right.}
\begin{tabular}{c@{\quad}c@{\qquad}r@{$|$}l@{\qquad}r@{$|$}l@{\qquad}r@{$|$}l}
\hline\hline
\rule{0pt}{12pt}& & \multicolumn{2}{c}{Mean squared error\hspace*{1.2em}}&
\multicolumn{2}{c}{Variance\hspace*{1.7em}} & \multicolumn{2}{c}{Bias}\\
\rule{0pt}{12pt}
Target & $F_0$ & LIN & MLE & LIN & MLE & LIN & MLE
\\
\hline
\rule{0pt}{14pt}
$\ket{\mbox{GHZ}_4}$ & 0.8 & $1.488\times10^{-4}$ & $2.623\times10^{-4}$ &
$1.480\times10^{-4}$ & $1.148\times10^{-4}$ & $8.236\times10^{-7}$ & $1.475\times10^{-4}$\\
\rule{0pt}{14pt}
$\ket{\mbox{GHZ}_4}$ & 1.0 & $0.0$
& $1.904\times10^{-6}$ & $0.0$ & $1.270\times10^{-7}$ &
$0.0$
& $1.777\times10^{-6}$\\
\rule{0pt}{14pt}
$\ket{\mbox{W}_4}$ & 1.0 & $2.323\times10^{-4}$ & $2.642\times10^{-6}$ &
$2.316\times10^{-4}$ & $1.792\times10^{-7}$ & $6.663\times10^{-7}$ & $2.463\times10^{-6}$\\
\rule{0pt}{18pt}
$\ket{\phi}$ & 0.981 & $3.082\times10^{-4}$ & $6.681\times10^{-4}$ &
$3.078\times10^{-4}$ & $6.449\times10^{-5}$ & $3.812\times10^{-7}$ & $6.036\times10^{-4}$\\
\rule{0pt}{14pt}
$\ket{\phi}$ & 0.016 & $3.044\times10^{-4}$ & $2.937\times10^{-4}$ &
$3.043\times10^{-4}$ & $7.704\times10^{-5}$ & $7.918\times10^{-8}$ & $2.166\times10^{-4}$\\
\rule{0pt}{14pt}
$\ket{\phi}$ & 0.8 & $3.472\times10^{-4}$ & $5.649\times10^{-4}$ &
$3.464\times10^{-4}$ & $1.418\times10^{-4}$ & $7.934\times10^{-7}$ & $4.231\times10^{-4}$\\
\rule{0pt}{18pt}
$\ket{\gamma}$ & 0.8 & $2.982\times10^{-4}$ & $5.614\times10^{-4}$ &
$2.982\times10^{-4}$ & $1.568\times10^{-4}$ & $9.143\times10^{-10}$ & $4.046\times10^{-4}$\\
\rule{0pt}{14pt}
$\ket{\tau}$ & 0.8 & $3.410\times10^{-4}$ & $5.087\times10^{-4}$ &
$3.410\times10^{-4}$ & $1.320\times10^{-4}$ & $3.371\times10^{-9}$ & $3.767\times10^{-4}$\\
\rule{0pt}{14pt}
$\ket{\theta}$ & 0.8 & $3.758\times10^{-4}$ & $5.859\times10^{-4}$ &
$3.758\times10^{-4}$ & $1.562\times10^{-4}$ & $4.369\times10^{-8}$ & $4.297\times10^{-4}$
\\[1ex]
\hline\hline
\end{tabular}
\end{table*}%

In all the examples above, LIN suffers from the problem of having unphysical
fidelity values.
Note that any attempt to fix these larger-than-1 or smaller-than-0 $F$ values
from LIN, \textsl{e.g.}, set all values ${>1}$ to be equal to $1$, will bias the
originally unbiased LIN estimator.
In all cases, LIN indeed gives a \emph{mean} fidelity closer to the true value
(in fact it should be exactly equal, if not for the fluctuations from only
$500$ runs) than MLE but, apart from the singular case of the GHZ state, has
a significantly larger spread than the MLE values, suggesting a possibly
larger, or at least comparable, MSE.
As mentioned in the previous section, for a single run of the experiment, what
matters is not the mean value over many repeated runs, but the MSE; a large
variance means that one has a high chance of winding up rather far away from
the true value.
Thus a meaningful comparison between MLE and LIN requires us to look at the
MSE for a variety of states.

Table \ref{tbl:MSE} presents the results of such a comparison.
The first six rows of the table give the variance, bias and MSE values
for the examples described above; the remaining rows give the values for
a further three randomly chosen target states, with the true
state in each case being the target state with added white noise to yield a
true target fidelity of $0.8$.
In all cases, LIN has zero bias (by construction), with the small deviations
indicated in the table attributable to finite statistics;
however, it has a variance comparable to that from the MLE in every single
case.
For the randomly chosen states, the MSE values from LIN are slightly smaller
than those from MLE, but again, MLE is not designed to reduce the
MSE (neither bias nor variance), and sacrificing physicality in LIN is a very
high price to pay for the occasional small reduction of the MSE.

\section{Conclusion}
The method of linear inversion may appear simple and intuitive---after all,
relative frequencies are asymptotically the correct frequentist's
interpretation of probabilities.
However, many difficulties arise, both in the construction of $\rhoLIN$
(having to put in ad-hoc fixes for internal consistency), as well as in the
interpretation of the estimated quantities (if one can compute them at all
from $\rhoLIN$, and having to deal with parameter values of no physical
meaning even when computation is possible).
Both these problems disappear when insisting on physical estimators.
This requirement is natural in all cases, since the true state must be
physical, not just asymptotically so. Even if one is willing to deal with the
problems of having a nonphysical estimator, reducing the MSE, not just the
bias, is the key, and the reduction had better be substantial enough for one
to want to cope with the nonphysicality issues.

To end the discussion, we remind the reader of a point briefly mentioned in
the opening paragraphs:
If one is interested in estimating only a single parameter, a direct
estimation from the data without first going through an estimator for the
state generally works better.
Thus, our assessment here of the LIN versus MLE is also lacking in that we
should compare their performance for a variety of parameters computed from
$\rhoLIN$ and $\rhoMLE$, for one would not report these estimators unless the
interest is in a few different parameters computed from the state.
Yet, given that LIN already does poorly in terms of target fidelity,
it is hardly necessary to provide more evidence against the use of LIN by
exploring other parameters.
Furthermore, a more comprehensive summary of the data is provided not by the
\emph{point} estimators $\rhoLIN$ and $\rhoMLE$, but by regions of estimators,
which can circumvent some of the problems highlighted here; see
\cite{ConfRegions} for confidence regions, and \cite{NJPpaper} for credible
regions.

\vspace*{1\baselineskip}

\acknowledgments
We thank Yong Siah Teo and David Nott for stimulating discussions.
The Centre for Quantum Technologies is a Research Centre of Excellence funded
by the Ministry of Education and the National Research Foundation of
Singapore.


\begin{thebibliography}{00}
\bibitem{note:FvsFsquare}
This $F$ is actually the square of the fidelity, if a close analog of the
fidelity of two probability distributions is wanted.
We here adopt the convention of Ref.~\cite{\tpaper} and use the term
``fidelity'' for $F$ thus defined.
\bibitem{note:data-split}
Note that splitting up a single set of data into very many smaller sets and
hence very many estimates does not help as the smaller sample sizes will
result in a larger variance.
\bibitem{note:irony}
What is particularly ironic is Jaynes's point that, for his example,
the estimator that minimizes the MSE actually requires one to ``double the
bias'' by dividing the variance by $N+1$ (for sample size $N$) instead of the
usual unbiased correction of $N-1$, as in Ref.~[37] of Ref.~\cite{\tpaper}.
\bibitem{note:GHZsingular}
If the GHZ state is the target, then the target fidelity involves the
expectation values of $[XXXX]$, $[YYYY]$, $[ZZZZ]$ as well as $[XXYY]$ and
$[ZZ\mathvisiblespace\mathvisiblespace]$ and the observables obtained by
permuting the four qubits.
The GHZ state is an eigenstate of every one of these $15$ observables.
Hence, there are no statistical fluctuations when their expectation values
are estimated from the data obtained from the product Pauli POM.
\end{thebibliography}
\end{document}